# Sensitivity Analysis for Causal Decomposition Analysis: Assessing Robustness Toward Omitted Variable Bias

Soojin Park[1]   Suyeon Kang[2]

Chioun Lee[3]   Shujie Ma[2]

[1]School of Education, University of California, Riverside

[2]Statistics, University of California, Riverside

[3]Sociology, University of California, Riverside

Abstract

A key objective of decomposition analysis is to identify a factor (the 'mediator') contributing to disparities in an outcome between social groups. In decomposition analysis, a scholarly interest often centers on estimating how much the disparity (e.g., health disparities between Black women and White men) would be reduced/remain if we set the mediator (e.g., education) distribution of one social group equal to another. However, causally identifying disparity reduction and remaining depends on the no omitted mediator-outcome confounding assumption, which is not empirically testable. Therefore, we propose a set of sensitivity analyses to assess the robustness of disparity reduction to possible unobserved confounding. We provide sensitivity analysis techniques based on regression coefficients and $R^2$ values. The proposed techniques are flexible to address unobserved confounding measured before and after the group status. In addition, $R^2$-based sensitivity analysis offers a straightforward interpretation of sensitivity parameters and a standard way to report the robustness of research findings. Although we introduce sensitivity analysis techniques in the context of decomposition analysis, they can be utilized in any mediation setting based on interventional indirect effects.

Sensitivity Analysis for Causal Decomposition Analysis: Assessing Robustness Toward

Omitted Variable Bias

## 1. Introduction

Decomposition analysis aims to identify factors that may close the observed gap in social, psychological, behavioral, or health outcomes between groups defined by social-demographic factors, such as gender/sex, race/ethnicity, socioeconomic status (SES), etc. Such factors are called 'mediators' because they are believed to lie between the exposure (one's social position) and the outcomes. Traditional decomposition analysis based on the difference-in-coefficients approach (Olkin & Finn, 1995; Freedman & Schatzkin, 1992) provides a straightforward way to estimate the degree to which the observed disparity is reduced or remains after controlling for the mediator. However, the traditional method is limited to a specific statistical model that cannot be readily generalizable to discrete mediators or outcomes and nonlinear relationships (Imai, Keele, & Tingley, 2010). Recently, several researchers developed and applied decomposition analysis within the counterfactual framework of causal inference, namely 'causal decomposition analysis', that overcomes the limitations of traditional decomposition analysis. VanderWeele and Hernán (2012) and Jackson and VanderWeele (2018) conceptualized decomposition analysis within the counterfactual framework, and many articles have appeared on this topic in the past five years (e.g., Jackson, 2017, 2018, 2020; Nguyen, Schmid, & Stuart, 2020; Park, Qin, & Lee, 2020; Lundberg, 2020).

A central goal of causal decomposition analysis is to estimate the degree to which an observed disparity would be reduced or remain if we equalize the mediator distribution across social groups. For example, how much would health disparities decrease if we equalize the education level between Black women and White men? The causal identification of the disparity reduction and remaining hinges on the strong assumption of no unobserved confounding in the mediator-outcome relationship. One way to assess the robustness of findings against possible violations of this assumption is to conduct a sensitivity analysis, yet few sensitivity analysis techniques are available in the causal decomposition framework. Previously, Park et al. (2020) developed a preliminary sensitivity analysis that assesses the robustness of disparity reduction and remaining estimates. However, Park and colleagues' method is restricted to a certain setting that requires conditional independence between unobserved and observed intermediate confounders (i.e., the effects of social groups confounding the mediator-outcome relationship). Another limitation is that the interpretation of sensitivity parameter is not straightforward since the prevalence difference in an unobserved confounder (e.g., being discriminated), comparing individuals in different groups, is conditioned on the mediator (e.g., education), which is a descendant of (a variable affected by) the group status.

In this study, we propose a set of sensitivity analyses for causal decomposition analysis, consisting of sensitivity parameters that are easy to interpret without making the restrictive assumption. After introducing our motivating example using data from Midlife Development in the U.S. (MIDUS) (Section 2), we review literature on causal decomposition analysis as a statistical framework that identifies contributing factors to disparities (Section 3). This review highlights how causal decomposition analysis differs from causal mediation analysis based on natural direct and indirect effects (Pearl, 2012; Robins, 2003). As a result, we show that sensitivity analysis developed for natural indirect effects can assess the robustness of disparity reduction when no intermediate confounders exist, which is unlikely in many studies that investigate contributing factors to disparities.

Therefore, we develop a set of sensitivity analyses for disparity reduction and remaining that incorporate 'observed' intermediate confounders. We begin by deriving general bias formulas for disparity reduction and remaining which are the basis of our proposed sensitivity analyses (Section 4). Since general bias formulas do not rely on any statistical models, they apply to various situations, including linear and nonlinear relationships as well as different types of mediator, outcome, or omitted confounder variables. We show that same bias formulas apply for 'unobserved' confounding measured before and after the group status. Second, we provide simplified bias formulas given linear models specified for the outcome and the unobserved confounder (Section 5). The simplified bias formulas offer a sensitivity analysis that is straightforward to use if the linearity assumption is met. Lastly, we reparameterize the regression-based sensitivity analysis to $R^2$ values (the proportion of variance explained) by extending Cinelli and Hazlett (2020) (Section 6). A critical advantage of this reparameterization to $R^2$ values is that we can estimate the correct standard errors with a varying amount of unobserved confounding. Another advantage is to provide a standard way of reporting the degree to which research findings are robust against the no unobserved confounding assumption. The standard way is referred to as the 'robustness value' by Cinelli and Hazlett (2020), which is the minimum strength of the confounder on the mediator and outcome, assuming an equal strength, to change research findings. The robustness value conveniently summarizes how sensitive the conclusions are to unobserved confounding.

In Section 7, we conclude with a discussion. Our sensitivity analysis is implemented in the 'causal.decomp' R package. Code to replicate all analyses can be found in Appendix F.

## 2. Running Example

To motivate the concepts and methods that we present, we rely on an epidemiological example; studies have consistently observed that racial and gender minorities, particularly Black women, show poorer cardiovascular health (CVH) than other race-gender groups. Socioeconomic status (SES), which is a fundamental cause and key determinant of access to resources, may operate via many mechanisms to affect multiple disease outcomes (Link & Phelan, 1995), including cardiovascular disease (Glymour, Clark, & Patton, 2014). Educational attainment plays a key role in explaining racial and gender disparities in health and education also affects other subsequent SES measures, such as income and wealth. Therefore, we hypothesize that the observed disparity in CVH between race-gender groups would decrease if we equalized the education levels between the groups. Causal

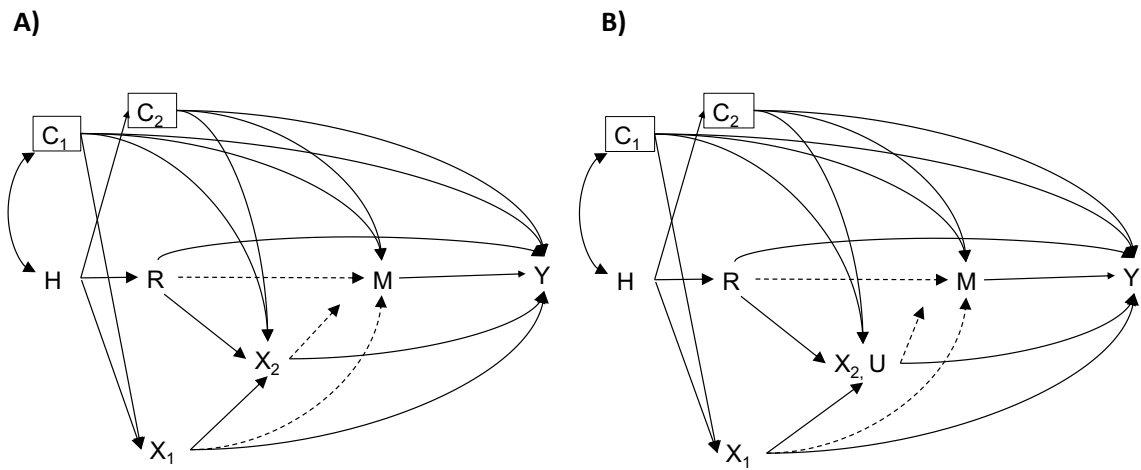

*Figure 1.* Directed acyclic graph (Pearl, 2001): A) When no unobserved confounder exists and B) When unobserved confounder $U$ exists.

Note. 1) Diagram represents the relationship between race and gender intersectional status $R$, cardiovascular health $Y$, and education $M$, as well as history $H$, age $C_1$, genetic vulnerability $C_2$, childhood SES $X_1$, and childhood abuse $X_2$.
2) Placing a box around the conditioning variables implies that a disparity is considered within levels of these variables.
3) Dotted lines indicate intervening on $M$ given baseline covariates.

decomposition analysis can be used to test this hypothesis.

From the hypothesis, we define four social groups: White men ($R = 0$), Black women ($R = 1$), Black men ($R = 2$), and White women ($R = 3$); the mediator is education ($M$); the outcome is CVH ($Y$). One concern is that education status is not randomized, and the relationship between education and later CVH could be confounded by various life course factors. Therefore, based on previous literature (Winkleby, Jatulis, Frank, & Fortmann, 1992; Suglia et al., 2018), we identified possible confounders, such as age ($C_1$), genetic vulnerability ($C_2$), and socioeconomic status (SES, $X_1$) and adverse experience in childhood ($X_2$).

We encodes our understanding of data-generation process involving these variables in Figure 1(A). We assume racial and gendered disparities in CVH arise through childhood abuse ($X_2$) and education ($M$). We also assume that these disparities could arise through historical processes that include racism and sexism (Kaufman, 2008). For example, due to historical processes, Blacks are more likely than Whites to be born into a family with low childhood SES ($X_1$) and suffer from a particular genetic vulnerabilities ($C_2$, parental history of cardiovascular and metabolic diseases). Age ($C_1$) also interacts with the historical process and in turn affects all other variables.

However, one may argue that the relationship between education and later CVH could still be confounded by unobserved confounders (e.g. discrimination based on race, gender, and social class) even after controlling for the observed variables. Figure 1 (B) shows a scenario in which an unobserved variable $U$ confounds the relationship between the mediator and the outcome. Depending on a kind of unobserved variable, it could be measured before, concurrently with, or after the group status. For example, if perceived discrimination were the unobserved confounder, $U$ would be measured after the group status, as shown in Figure 1(B). As another example, if an unknown genetic factor that affects later education and CVH were the unobserved confounder, $U$ would be measured before the group status.

How can we validate our conclusions on disparity reduction and remaining to the possible omission of unobserved variables? In Section 4, we present a set of sensitivity analyses that applied researchers can use to validate their findings to possible omitted variables. To proceed, investigators should further clarify baseline covariates and intermediate confounders, in addition to identifying the group status, the mediator (education), and the outcome (CVH). Baseline covariates characterize demographics through which CVH or education differences are considered equitable (Jackson, 2020), which are age or genetic vulnerability in our example. Intermediate confounders represent the effects of race-gender status that confound education and CVH, which are childhood SES and abuse in the example. We use the following notation to represent baseline covariates and intermediate confounders. Intermediate confounders are denoted as $\mathbf{X} = (X_1, X_2)$ where $\mathbf{x} \in \mathcal{X}$; baseline covariates are denoted as $\mathbf{C} = (C_1, C_2)$ where $\mathbf{c} \in \mathcal{C}$.

## 3. Review and Implications on Existing Sensitivity Analyses

This section provides an overview of causal decomposition analysis based on interventional analogues of natural direct and indirect effects (interventional effects, VanderWeele & Vansteelandt, 2014; VanderWeele, Vansteelandt, & Robins, 2014). We then compare this approach to causal mediation analysis based on natural direct and indirect effects. Finally, we show the conditions in which sensitivity analyses developed for natural indirect effects can be used to assess the sensitivity of disparity reduction.

### 3.1. Causal Decomposition Analysis

Causal decomposition analysis does not contribute to the argument of whether socially defined characteristics such as race and gender can be given a causal interpretation (Jackson & VanderWeele, 2018). Rather, it focuses on estimating a causal effect of potentially manipulable factors (mediators) in reducing the observed disparity. Therefore, causal decomposition analysis aims to estimate how much the disparity would be reduced by intervening to equalize the mediator distribution between social groups. In the motivating example, this intervention implies increasing the Black women's education to the level of White men among those with the same baseline covariates. To illustrate, we use the example of comparing Black women (comparison group) and White men (reference group). Statistically, it requires imputing each Black women's mediator with a randomly drawn value from the mediator distribution of White men among those with the same level of baseline covariates. A random draw from the mediator distribution of the reference group given $\mathbf{C} = \mathbf{c}$ is denoted as $G_{m|\mathbf{c}}(0)$.

**Definitions.**    Using this notation, disparity reduction is defined as, given baseline covariates, the difference between the average CVH of a comparison group and the average counterfactual CVH if their education level was the same as the reference group among those with the same baseline covariates level. Formally,

$$\delta(r) \equiv E[Y_i|R_i = r, \mathbf{c}] - E[Y_i(G_{m|\mathbf{c}}(0))|R_i = r, \mathbf{c}] \ \ for \ r \in \{1,2,3\} \ \text{ and } \mathbf{c} \in \mathcal{C}. \quad (1)$$

Likewise, the disparity remaining is defined as, given baseline covariates, the difference between the average counterfactual CVH of a comparison group after the hypothetical intervention and the average CVH outcome of the reference group. Formally,

$$\zeta(0) \equiv E[Y_i(G_{m|\mathbf{c}}(0))|R_i = r, \mathbf{c}] - E[Y_i|R_i = 0, \mathbf{c}] \ \ for \ r \in \{1,2,3\} \ \text{ and } \mathbf{c} \in \mathcal{C}. \quad (2)$$

The observed disparity is equal to the sum of disparity reduction and disparity remaining due to hypothetically intervening on the mediator as $\tau(r,0) = \delta(r) + \zeta(0)$ for $r \in \{1,2,3\}$.

Disparity reduction and disparity remaining correspond to interventional indirect and direct effects, respectively. The term 'interventional' implies that these definitions involve interventions (Didelez, Dawid, & Geneletti, 2012), although hypothetical, of mediator that may inform actual interventions. For instance, a significant disparity reduction in CVH due to intervening on the education level can inform actual interventions of promoting access to college for minorities.

**Identification Assumptions and Results.** Defining disparity reduction and remaining requires an unobservable quantity–i.e., $E[Y_i(G_{m|c}(0))|R_i = r, \mathbf{c}]$, which is an inherent problem of causal inference (Holland, 1986). Therefore, we invoke the following assumptions to identify disparity reduction and disparity remaining.

- **A1: Conditional Independence**: $Y_i(m) \perp M_i|R_i = r, \mathbf{X}_i = \mathbf{x}, \mathbf{C}_i = \mathbf{c}$, for all $r \in \{0,1,2,3\}, \mathbf{x} \in \mathcal{X}, m \in \mathcal{M}$ and $\mathbf{c} \in \mathcal{C}$. There is no omitted confounding in the mediator-outcome relationship given the race-gender status and measured confounders ($\mathbf{X}$ and $\mathbf{C}$).

- **A2: Positivity**: $0 < P(M_i = m|R_i = r, \mathbf{x}, \mathbf{c})$ or all $r \in \{0,1,2,3\}, \mathbf{x} \in \mathcal{X}, m \in \mathcal{M}$

and $\mathbf{c} \in \mathcal{C}$. The conditional probability of $m$ for a comparison group is positive given measured confounders.

- **A3: Consistency**: if $M_i = m$ then $Y_i = Y_i(m)$ for all $m \in \mathcal{M}$, where $Y_i(m)$ is the potential value of the outcome for individual $i$ under $M = m$. Consistency states that the observed outcome under the exposure history is the same as the potential outcome after setting the exposure to that level.

These three assumptions are all strong, and the plausibility of the assumptions depends on the context of the study. This study aims to address the possible violation of conditional independence.

Given the assumptions, the disparity reduction and remaining are nonparametrically identified as

$$\begin{aligned} \delta(r) =& E[Y_i | R_i = r, \mathbf{c}] - \sum_{\mathbf{x},m} E[Y_i | R_i = r, \mathbf{x}, m, \mathbf{c}] P(\mathbf{x} | R_i = r, \mathbf{c}) P(m | R_i = 0, \mathbf{c}) \text{ and} \\ \zeta(0) =& \sum_{\mathbf{x},m} E[Y_i | R_i = r, \mathbf{x}, m, \mathbf{c}] P(\mathbf{x} | R_i = r, \mathbf{c}) P(m | R_i = 0, \mathbf{c}) - E[Y_i | R_i = 0, \mathbf{c}], \end{aligned} \tag{3}$$

where $r \in \{1, 2, 3\}, x \in \mathcal{X}, m \in \mathcal{M}_1$, and $\mathbf{c} \in \mathcal{C}$.

**Estimation.** If the assumptions hold, there are many ways to estimate disparity reduction and disparity remaining including regression-based estimators (Jackson & VanderWeele, 2018; Park, Kang, & Lee, 2021), weighting-based estimators (Jackson, 2020), and imputation-based estimators (Park et al., 2020; Sudharsanan & Bijlsma, 2021). Although limited due to its modeling assumptions, regression-based methods are perhaps the most straightforward way to estimate disparity reduction and remaining. Consider the following models fitted to the intermediate confounders, mediator, and outcome as,

$$X_i = \gamma + \sum_r \gamma_r I_i(r) + \gamma_c \mathbf{C}_i + \epsilon_{x,i}, \tag{4}$$

$$M_i = \alpha + \sum_r \alpha_r I_i(r) + \alpha_c \mathbf{C}_i + \epsilon_{m,i}, \quad \text{and} \tag{5}$$

$$Y_i = \beta + \sum_r \beta_r I_i(r) + \beta_x \mathbf{X}_i + \beta_m M_i + \beta_c \mathbf{C}_i + \epsilon_{y,i}, \tag{6}$$

where $r \in \{1, 2, 3\}$ and $\epsilon_{m,i}$ and $\epsilon_{y,i}$ follow a normal distribution. Under these models, disparity reduction is estimated as $\hat{\delta}(r) = \hat{\alpha}_r \hat{\beta}_m$; disparity remaining is estimated as $\hat{\zeta}(0) = \hat{\beta}_r + \hat{\beta}_x \hat{\gamma}_r$. Note that this regression-based estimator hinges on the functional form assumptions made in the models shown in equations (4), (5), and (6). If the interaction effect exists between the social group and the mediator, one should include the interaction term in the outcome model as

$$Y_i = \beta + \sum_r \beta_r I_i(r) + \beta_x \mathbf{X}_i + \beta_m M_i + \sum_r \beta_{rm} I_i(r) M_i + \beta_c \mathbf{C}_i + \epsilon_{y,i}. \tag{7}$$

Then, the disparity reduction is estimated as $\hat{\delta}(r) = \hat{\alpha}_r(\hat{\beta}_m + \hat{\beta}_{rm})$; disparity remaining is estimated as $\hat{\zeta}(0) = \hat{\beta}_r + \hat{\beta}_x \hat{\gamma}_r + \hat{\beta}_{rm}(\hat{\alpha} + \hat{\alpha}_c \hat{E}[\mathbf{C}_i])$. A proof is given in Park et al. (2021) and is, thus, omitted here. As far as investigators are willing to make the modeling assumptions along with A1-A3, these estimators provide a valid estimate of disparity reduction and disparity remaining.

## 3.2. Comparison to Natural Direct and Indirect Effects

In this section, we compare disparity reduction and remaining defined in equations (1) and (2) with the related definitions in the causal mediation literature. First, we argue that defining natural indirect effects is not straightforward in disparities research. In the example, the natural indirect effect defined between Black women and White men in the motivating example is the expected difference, comparing each Black woman's actual and potential CVH after setting her education level to a value that would have naturally been observed had she been born a White man. Given that gender and race are essentially non-modifiable, it is somewhat strange to consider fixing each individual's mediator to a value that would have naturally been observed had the individual been born a White man (Jackson & VanderWeele, 2019). In contrast, interpreting disparity reduction defined in equation (1) is straightforward because interventional effects allow defining disparity reduction at the group level. Disparity reduction compares the average Black woman's CVH to the average counterfactual CVH outcome of Black women after equalizing their education level to that of White men as a group.

Next, we compare the conditional independence assumption (A1) with the related assumptions in the causal mediation literature. We don't compare assumptions regarding the exposure (treatment ignorability). Causal decomposition analysis does not attempt to estimate the causal effect of social groups (race and gender) and thus does not make any assumption regarding the exposure (group status). Pearl (2001) made the following assumption to identify natural indirect and direct effects: $Y_i(r, m) \perp M_i(r')|\mathbf{C}_i = c$, which implies 1) no unobserved pre-exposure confounders (i.e., variables measured before the exposure that confound the mediator-outcome relationship) given the race-gender status and baseline covariates and 2) no intermediate confounders. The first assumption might be met in some disparities research since there are hardly any variables that can be measured before race and gender status. However, the second assumption is unlikely to be met in many disparities research settings, since a myriad of life-course factors contribute to disparities in education and CVH.

Alternatively, Robins (2003) made the following assumption: $Y_i(r, m) \perp M_i(r)|R_i = r, \mathbf{X}_i = x, \mathbf{C}_i = c$, which allows intermediate confounders $\mathbf{X}$. This is an important advantage considering that the assumption of no intermediate confounders is unlikely to be met in many disparities research settings. However, this relaxed assumption comes with the cost of requiring no interactions between the exposure and the mediator at

the individual level. Unfortunately, assuming no interaction effects between the group status and the mediator is also a strong assumption that is unlikely to be met in many studies. For example, prior studies documented diminished returns of human capital as a result of discrimination for minorities (Assari, 2020). In comparison, interventional effects require neither the no-interaction assumption nor the no-intermediate-confounders assumption.

### 3.3 Implications on Existing Sensitivity Analyses

**Under No Intermediate Confounders.** If no intermediate confounders exist, interventional direct and indirect effects coincide with natural direct and indirect effects, respectively. Although it may be a rare situation, suppose that the intermediate confounders (e.g., childhood SES and abuse) do not exist. Then, one can use a nonparametric sensitivity analysis (VanderWeele, 2010; Hong, Qin, & Yang, 2018) or parametric sensitivity analysis (Imai et al., 2010) developed for natural direct and indirect effects to assess the robustness of estimated disparity reduction and remaining to possible unobserved pre-exposure confounding.

**Under No Interaction Between $R$ and $M$.** Now, we consider a situation where intermediate confounders exist, but no interaction exists between social groups and the mediator. This situation is perhaps plausible in some studies that examine disparities depending on a type of mediator. Then, the estimator of disparity reduction $\alpha_r \beta_m$ is consistent for the natural indirect effect given equations (5) and (6). However, sensitivity analysis developed for natural direct and indirect effects to possible violations of unobserved intermediate confounding (e.g., Imai & Yamamoto, 2013; VanderWeele & Chiba, 2014) is not appropriate for assessing the robustness of estimated disparity reduction and remaining. For instance, a sensitivity analysis developed by Imai and Yamamoto (2013) addresses the bias due to incorporating interaction effects given that all intermediate confounders are observed. VanderWeele and Chiba (2014) do not assume that all intermediate confounders are observed; however, the bias formulas do not incorporate existing intermediate confounders.

In conclusion, a new sensitivity analysis is necessary for disparity reduction and remaining that incorporates existing intermediate confounders and possible interaction effects.

## 3.4. An Application to MIDUS

We estimate disparity reduction and remaining with varying assumptions. Table 1 shows the estimated quantities of interest between Black women and White men. Other comparisons are available but we only present the results between Black women and White men for simplicity. The initial disparity between Black women and White men is negative (-0.965) and the 95% confidence interval is bounded away from zero, meaning that Black women have significantly worse CVH than White men after controlling for age and genetic vulnerability.

First, we estimate disparity reduction and remaining, assuming no interaction between education and race-gender status (first column in Table 1). The estimates are consistent for natural direct and indirect effects defined by Robins (2003). Disparity reduction is negative (-0.401) and is significant at the 95% confidence level. The initial disparity would decrease by 41.6% if Black women's education level were the same as White men's among those with the same age and genetic vulnerability level.

We compare this result after relaxing the no interaction assumption (third column), which is consistent with disparity reduction and remaining defined in equations (3). The difference in disparity reduction and remaining estimates is small. Disparity reduction changes from -0.401 (first column) to -0.360 (third column); disparity remaining changes from -0.564 (first column) to -0.604 (third column). This slight change after relaxing the no interaction assumption implies that there is little evidence for the presence of the interaction effect.

Table 1

*Estimates of the disparity reduction and disparity remaining*

|  | Estimate (95% CI) | | |
|---|---|---|---|
| $R - M$ Interaction | No | Yes | Yes |
| Intermediate Confounders | Yes | No | Yes |
| Initial disparity ($\tau(1,0)$) | -0.965 | -0.965 | -0.965 |
| (95% CI) | (-1.259, -0.662) | (-1.257, -0.675) | ( -1.238, -0.658) |
| Disparity remaining ($\zeta(0)$) | -0.564 | -0.480 | -0.604 |
| (95% CI) | (-0.863, -0.278) | (-0.791, -0.143) | (-0.868, -0.187) |
| Disparity reduction ($\delta(1)$) | -0.401 | -0.485 | -0.360 |
| (95% CI) | (-0.561, -0.239 ) | (-0.727, -0.261) | (-0.712, -0.219) |
| % reduction | 41.6% | 50.2% | 37.3% |

Note. Black women: $R = 1$, White men: $R = 0$.

Lastly, we examine the result after relaxing the no interaction assumption but without intermediate confounders (second column), which is consistent with the natural indirect effect defined by Pearl (2001). Not controlling for intermediate confounders results in an over- or under-estimation of disparity reduction and remaining when compared to controlling for intermediate confounders (third column). After including observed intermediate confounders, disparity reduction changes from -0.485 (second column) to -0.360 (third column); disparity remaining changes from -0.480 (second column) to -0.604 (third column). This result suggests over-estimation of disparity reduction due to education when intermediate confounders were not accounted for. In the next section, we present a sensitivity analysis to possible unobserved confounding that incorporate observed intermediate confounders and the interaction effect.

## 4. General Bias Formulas

We begin by calculating the bias for disparity reduction and remaining when the unobserved confounder $U$ exists, on which the two sensitivity analyses that we propose in Sections 5 and 6 are based. We first derive bias formulas when intermediate unobserved confounding exist and later show that the same bias formulas apply, with an additional assumption, to pre-exposure unobserved confounding.

The conditions required to calculate the bias are as follows: 1) no omitted confounding exists in the mediator and outcome relationship given group status ($r$), observed confounders ($\mathbf{x}, \mathbf{c}$), and the unobserved confounder ($u$), as

$M_i \perp Y_i(m)|R_i = r, \mathbf{X}_i = \mathbf{x}, \mathbf{C}_i = \mathbf{c}, U_i = u$ and 2) the unobserved confounder ($U$) is an effect of group status ($R$) and thereby is considered as an intermediate unobserved confounder as shown in Figure 1(B). An example of intermediate unobserved confounder includes discrimination, which is an effect of group status and adversely affects education and CVH.

Suppose that we only had observed data. Then, researchers would often estimate the disparity reduction using observed data as,

$\delta_{res}(r) = E[Y_i|R_i = r, \mathbf{c}] - \sum_{\mathbf{x},m} E[Y_i|R_i = r, \mathbf{x}, m, \mathbf{c}]P(\mathbf{X} = \mathbf{x}|R_i = r, \mathbf{c})P(M_i = m|R_i = 0, \mathbf{c})$, as equation (3). However, if the unobserved confounder $U$ exists, this expression will lead to a biased estimate of disparity reduction. The bias is therefore defined as the difference between this estimator using observed data and the true effect of disparity reduction as

$$
\begin{aligned}
bias(\delta(r)) =& \delta_{res}(r) - \delta(r) \\
=& E[Y_i|R_i = r, \mathbf{c}] - \sum_{\mathbf{x},m} E[Y_i|R_i = r, \mathbf{x}, m, \mathbf{c}]P(\mathbf{x}|R_i = r, \mathbf{c})P(m|R_i = 0, \mathbf{c}) \\
& - E[Y_i|R_i = r, \mathbf{c}] + \sum_{\mathbf{x},m,u} E[Y_i|R_i = r, \mathbf{x}, m, \mathbf{c}, u]P(\mathbf{x}, u|R_i = r, \mathbf{c})P(m|R_i = 0, \mathbf{c}).
\end{aligned}
$$

(8)

Note that we used $P(\mathbf{x}, u|R_i = r, \mathbf{c})$ to accommodate both cases: 1) when $U$ is measured before $\mathbf{X}$ and 2) when $U$ is measured after $\mathbf{X}$. The bias for disparity remaining is defined the same way ($bias(\zeta(0)) = \zeta_{res}(0) - \zeta(0)$).

Then, for a particular value of $U = u'$, the biases for disparity reduction and remaining are given by

$bias(\delta(r)) = \sum_{\mathbf{x},m,u} \{E[Y|r, \mathbf{x}, m, \mathbf{c}, u] - E[Y|r, \mathbf{x}, m, \mathbf{c}, u']\}\{P(u|r, \mathbf{x}, c) - P(u|r, \mathbf{x}, m, \mathbf{c})\}P(\mathbf{x}|r, \mathbf{c})P(m|R = 0, \mathbf{c}),$

$bias(\zeta(0)) = -\sum_{\mathbf{x},m,u} \{E[Y|r, \mathbf{x}, m, \mathbf{c}, u] - Y|r, \mathbf{x}, m, \mathbf{c}, u']\}\{P(u|r, \mathbf{x}, c) - P(u|r, \mathbf{x}, m, \mathbf{c})\}P(\mathbf{x}|r, \mathbf{c})P(m|R = 0, \mathbf{c})$

(9)

A proof is given in Appendix A. These general bias formulas do not require any assumptions regarding functional forms, variable types of the mediator, outcome, or unobserved confounders. While these bias formulas can be used in general settings, their applicability may be limited due to too many moving parts (i.e., sensitivity parameters). Therefore, we provide simplified bias formulas under linear models specified for the outcome and the unobserved confounder that are straightforward to use in Section 6. We will discuss the applicability of the general bias formulas later.

Note that the same bias formulas apply with pre-exposure unobserved confounding under the additional assumption of $R \perp U | \mathbf{C}$, where $U$ is pre-exposure unobserved confounder. The assumption states that the pre-exposure confounder and group status do not affect each other given baseline covariates. An example of such confounder would be an unknown genetic factor that is not related to race or gender, but affects later education and CVH. Given the pre-exposure confounder $U$, the bias for disparity reduction is defined as

$$bias(\delta(r)) = E[Y_i | R_i = r, \mathbf{c}] - \sum_{\mathbf{x}, m} E[Y_i | R_i = r, \mathbf{x}, m, \mathbf{c}] P(\mathbf{x} | R_i = r, \mathbf{c}) P(m | R_i = 0, \mathbf{c})$$

$$- E[Y_i | R_i = r, \mathbf{c}] + \sum_{\mathbf{x}, m, u} E[Y_i | R_i = r, \mathbf{x}, m, \mathbf{c}, u] P(\mathbf{x} | R_i = r, \mathbf{c}, u) P(m | R_i = 0, \mathbf{c}) P(u | \mathbf{c}).$$

This equation leads to the same expression shown in equation (8) since $P(\mathbf{x} | R_i = r, \mathbf{c}, u) P(u | \mathbf{c}) = P(\mathbf{x} | R_i = r, \mathbf{c}, u) P(u | R_i = r, \mathbf{c})$ due to the assumption of $R \perp U | \mathbf{C}$.

## 5. Sensitivity Analysis Using Regression Coefficients

Unlike the general bias formulas, the proposed sensitivity analyses in this study are developed under a particular statistical model that assumes linearity. However, we show that an extension is possible when linearity is violated.

### 5.1. Under the Linearity Assumption

The general bias formulas in equations (9) can be simplified under linear models specified for the outcome and the unobserved confounder as

$$E[Y_i|r, x, m, c, u] = \beta + \beta_r + \beta_x x + \beta_m m + \beta_c c + \beta_u u, \text{ and}$$
$$E[U_i|r, x, m, c] = \delta + \delta_r + \delta_x x + \delta_m m + \delta_c c, \tag{10}$$

for $r \in \{1, 2, 3\}$. We first assume the simplest models for the outcome and the unobserved confounder and discuss later how to relax the linearity assumption. Given equations (10), the nonparametric bias formulas for disparity reduction and remaining are considerably simplified as

$$bias(\delta(r)) = \alpha_r \delta_m \beta_u, \text{ and}$$
$$bias(\zeta(0)) = -\alpha_r \delta_m \beta_u \quad \text{for } r \in \{1, 2, 3\}, \tag{11}$$

where an unbiased estimate of $\alpha_r$ can be obtained by fitting the mediator model in equation (5). A proof is given in Appendix B.

We offer several remarks about these bias formulas. First, the bias for disparity reduction and remaining is the same except for the sign. This is because the initial disparity is an observed quantity conditional on specified covariates, hence no bias exists due to the unobserved mediator-outcome confounder.[1]

Second, the bias is zero when either $\beta_u$ or $\delta_m$ is zero, meaning that the unobserved confounder is not associated with the outcome or the mediator given observed confounders. The bias is also zero when the mediator does not differ by group given baseline covariates.

Third, the unobserved variable confounds the intermediate confounders-outcome and intermediate confounders-mediator relationships ($\mathbf{X} - Y$ and $\mathbf{X} - M$) in addition to the mediator-outcome relationship ($M - Y$). However, confounding relationships in the $\mathbf{X} - Y$ and $\mathbf{X} - M$ relationships do not contribute to the bias of disparity reduction and remaining. The only confounding path that matters for disparity reduction and remaining is $M \leftarrow U \rightarrow Y$. Intuitively, it makes sense since conditional independence (A1) only concerns unobserved confounding in the mediator-outcome relationship.

---

[1]
$$bias(\tau(1, 0)) = bias(\delta(r) + bias(\zeta(0))$$
$$= \alpha_r \delta_m \beta_u - \alpha_r \delta_m \beta_u = 0 \tag{12}$$

Finally, the intermediate confounder and outcome models specified in equations (10) imply that the effect of unobserved confounder ($U$) on the outcome ($Y$) is constant across social groups. However, it may be too restrictive to assume the constant effect of $U$ across social groups. For example, discrimination has a differential effect on CVH by race (Bey, Jesdale, Forrester, Person, & Kiefe, 2019). To address this differential impact of $U$ on $Y$, we can add the interaction effect between $u$ and $I(r)$ in the outcome model as $\sum_r \beta_{ru} I(r) u$. Then the bias for disparity reduction is given by $bias(\delta(r)) = \alpha_r \delta_m (\beta_u + \beta_{ru})$.

The bias formulas expressed in regression coefficients give us an intuitive idea of sensitivity analysis. The basic idea is to find a combination of $\beta_u$ and $\delta_m$ that will explain away the disparity reduction and remaining; change the significance of the effects. Given the combinations, researchers are required to determine whether the amount of confounding expressed in these sensitivity parameters is plausible or not. To do so, understanding the precise meaning of the sensitivity parameters is essential. One sensitivity parameter ($\beta_u$) is the difference in the outcome, comparing individuals who differ by one unit on the confounder $U$, after controlling for the group status, the mediator, and observed confounders. Another sensitivity parameter ($\delta_m$) is the difference in the unobserved confounder $U$, comparing individuals who differ by one unit on the mediator $M$, after controlling for race-gender status and observed confounders. We illustrate the use of this sensitivity analysis with the MIDUS example in Section 5.3.

## 5.2. When the Linearity Assumption Is Violated

When the linearity assumption made in equations (10) is violated, one can either 1) derive their bias formulas based on modified models for $Y$ and $U$ or 2) use the original bias formula shown in equation (9). We showed earlier how to address the differential impact of the unobserved confounder $U$ on the outcome $Y$ by group status. However, addressing another differential impact (e.g., the interaction between the unobserved confounder and the mediator) leads to a more complex form of bias formulas. In this case, researchers can directly use the nonparametric bias formulas. To use the nonparametric bias formulas, one should choose a reference value for $U = u'$ and specify $E[Y_i|r, \mathbf{x}, m, \mathbf{c}, u] - E[Y_i|r, \mathbf{x}, m, \mathbf{c}, u']$, which is the difference in the outcome among the comparison group $R = r$, comparing $U = u$ and $U = u'$, across strata of $\mathbf{x}, m,$ and $\mathbf{c}$. Also, one should specify $P(u|r, \mathbf{x}, c) - P(u|r, \mathbf{x}, m, \mathbf{c})$, which is the distribution of the unobserved confounder $U$ for the comparison group $R = r$, conditional on $\mathbf{X} = \mathbf{x}$ and $\mathbf{C} = \mathbf{c}$, compared with the distribution of the unobserved confounder $U$ conditional on $\mathbf{X} = \mathbf{x}$, $\mathbf{C} = \mathbf{c}$, and $M = m$. As mentioned earlier, using the original bias formulas is not entirely straightforward since specifying these values may be difficult. However, in some cases, it may be possible to draw these values based on substantive knowledge and use the bias formulas directly.

## 5.3. Illustration Using Regression-Based Sensitivity Analysis

This section uses the example provided in Section 3.4 to describe and interpret the sensitivity analysis using regression coefficients. In Section 3.4, we estimated the disparity reduction and disparity remaining between Black women and White men. We here focus on the case after assuming differential effects of education by group ($R - M$ interaction) and accounting for existing intermediate confounders. As shown in Table 2, the initial disparity would be reduced by about 37.3% if Black women's education level was the same as that of White men among those with the same age and genetic vulnerability. This result can be given a causal interpretation under the assumptions described in A1-A3. Suppose that

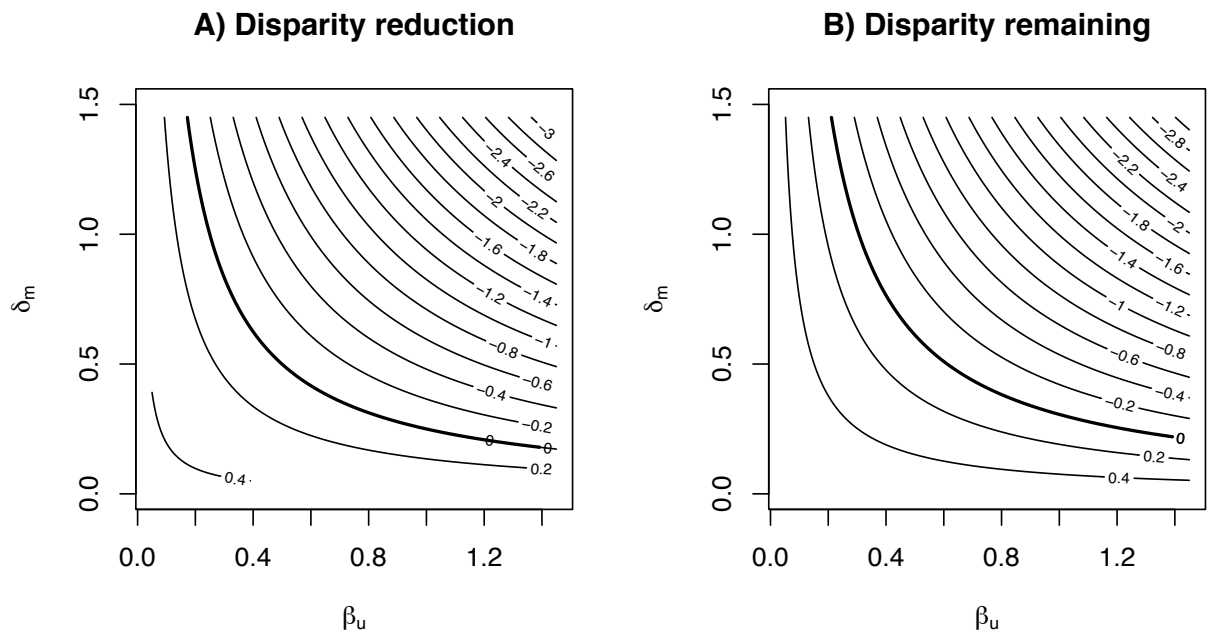

*Figure 2*. Sensitivity contour plots using regression coefficients.

Note. 1) Bold lines represents the points at which the estimates become zero. 2) Standard lines represent the points at which the estimates become the respective value (e.g., 0.4, 0.2, -0.2, -0.4).

conditional independence (A1) is violated because an unobserved variable, that is, whether each individual perceives to be discriminated or not, confounds education and CVH. How strongly does this unobserved confounder have to be to explain away the estimates?

In Figure 2, hypothetical values for $\beta_u$ and $\delta_m$ lie on the horizontal and vertical axis, respectively. The contour lines show the true disparity reduction and remaining values at hypothesized values of $\beta_u$ and $\delta_m$. In addition to the figure, we can also calculate the true disparity reduction and remaining given $\hat{\alpha} = -1.736$. Suppose that the difference in CVH between those who are discriminated and those not discriminated is, say, $\beta_u = 0.993$ (as strong as the effect of education for White men, which is the strongest among four race-gender groups) after controlling for the group status, the mediator, and observed confounders. Then to completely explain away the estimated disparity reduction ($\delta(1) = -0.360$), $\delta_m$ would need to be close to 0.208 (i.e., $-0.360 - (-1.736 \times 0.993 \times 0.208) \approx 0$). This $\delta_m$ value indicates that the probability difference of being discriminated is 20.8%, when comparing individuals with the same values for the group status and observed confounders (childhood SES and abuse, education, age, and genetic vulnerability), but differing by one unit on education. Given the same $\beta_u$, to completely explain away the estimated disparity remaining ($\zeta(1) = -0.604$), $\delta_m$ would need to be close to 0.350 (i.e., $-0.604 - (-1.736 \times 0.993 \times 0.350) \approx 0$). These $\delta_m$ values seem unlikely large given that we compare those who have the same group status, age, and genetic vulnerability, but differing by one unit in education (e.g., no schooling vs. graduating junior high school).

## 6. Sensitivity Analysis Using $R^2$ Values

The regression-based sensitivity analysis has often been used with a binary unobserved confounder, such as whether discriminated or not. It is perhaps because the interpretation of sensitivity parameters with a binary unobserved confounder is straightforward. Specifically, $\delta_m$ is the *prevalence difference* in the unobserved confounder $U$, comparing individuals with one unit increase in the mediator $M$, given controls; $\beta_u$ is the outcome difference between *two levels of the unobserved confounder*, given controls. However, the interpretation of sensitivity parameters is no longer straightforward if the unobserved confounder is continuous, such as discrimination intensity. With a continuous unobserved confounder, sensitivity parameters depend on the scale of the unobserved confounder. To see this, $\delta_m$ is the *difference in the unobserved confounder*, comparing individuals with a one-unit increase in the mediator $M$, give controls; $\beta_u$ is the outcome difference, comparing individuals with a *one-unit increase in the unobserved confounder*, given controls. If sensitivity parameters depend on the scale of the unobserved confounder, it is challenging to determine whether the confounding amount expressed by the sensitivity parameters is large or small even with the substantial knowledge of possible unobserved confounders (e.g., degree of discrimination). One solution is to standardize the unobserved confounder as shown in Section 5.3. Another solution is to reparameterize sensitivity parameters to $R^2$ values.

In this section, we re-express the simplified bias formulas shown in equations (11) using scale-free $R^2$ values. Using the $R^2$ values, we can obtain correct standard errors for disparity reduction and remaining with a varying amount of unobserved confounding. We also present a standard way to report the amount of confounding using the robustness values. Cinelli and Hazlett (2020) developed a way to express biases using $R^2$ values in the context of estimating a treatment effect, and we extend this reparameterization of $R^2$ values to our bias formulas for disparity reduction and remaining.

This reparameterization to $R^2$ values is based on a particular statistical model specified in equations (5) and (6) that assumes linearity. We later provide the extension that can be used when linearity is violated. Note that we rewrite equation (6) as below to differentiate outcome coefficients before and after including the unobserved confounder $U$.

$$E[Y_i|r, x, m, c, u] = \beta_{res} + \beta_{res,r} + \beta_{res,x}x + \beta_{res,m}m + \beta_{res,c}c \tag{13}$$

## 6.1. Under the Linearity Assumption

**Point Estimates.** Let the partial $R^2$ value of the unobserved confounder ($U$) with the outcome ($Y$) for the comparison group ($R = r$) given mediator ($M$) and observed confounders ($\mathbf{X}$ and $\mathbf{C}$) be denoted as $R^2_{Y \sim U|r, \mathbf{X}, M, \mathbf{C}}$; and the partial $R^2$ value of the unobserved confounder ($U$) with the mediator ($M$) for the comparison group ($R = r$) given observed confounders ($\mathbf{X}$ and $\mathbf{C}$) be denoted as $R^2_{M \sim U|r, \mathbf{X}, \mathbf{C}}$. Then, the absolute value of bias for disparity reduction and remaining is expressed as

$$|bias(\delta(r))| = |bias(\zeta(0))| = |\alpha_r| \sqrt{Var(\hat{\beta}_{res,m})} \sqrt{\frac{R^2_{Y \sim U|r, \mathbf{X}, M, \mathbf{C}} \times R^2_{M \sim U|r, \mathbf{X}, \mathbf{C}}}{1 - R^2_{M \sim U|r, \mathbf{X}, \mathbf{C}}} df}, \tag{14}$$

for $r \in \{1, 2, 3\}$. We can obtain an unbiased estimate of $\alpha_r$ by fitting the mediator model shown in equation (5); $Var(\hat{\beta}_{res,m})$ is obtained from the sample variance of $\hat{\beta}_{res,m}$ and $df$ is obtained from the degrees of freedom of the outcome model shown in equation (13). A proof is given in Appendix C.

**Standard Errors.** Some investigators might also be interested in quantifying the amount of confounding that would change the significance of the effects. The standard error for disparity reduction for $R = r$ can be calculated approximately using the Delta method (Sobel, 1982) as

$$Var(\hat{\delta}(r)) \approx \alpha_r^2 Var(\hat{\beta}_m) + \beta_m^2 Var(\hat{\alpha}_r), \tag{15}$$

where $Var(\hat{\beta}_m) = Var(\hat{\beta}_{res,m})(\frac{1-R^2_{Y \sim U|r,\mathbf{X},M,\mathbf{C}}}{1-R^2_{M \sim U|r,\mathbf{X},\mathbf{C}}} \frac{df}{df-1})$ and

$\beta_m = \beta_{res,m} - \sqrt{Var(\hat{\beta}_{res,m})} \sqrt{\frac{R^2_{Y \sim U|r,\mathbf{X},M,\mathbf{C}} \times R^2_{M \sim U|r,\mathbf{X},\mathbf{C}}}{1-R^2_{M \sim U|r,\mathbf{X},\mathbf{C}}} df}$, assuming $\beta_{res,m}$ is positive. If $\beta_{res,m}$

is negative, $\beta_m = \beta_{res,m} + \sqrt{Var(\hat{\beta}_{res,m})} \sqrt{\frac{R^2_{Y \sim U|r,\mathbf{X},M,\mathbf{C}} \times R^2_{M \sim U|r,\mathbf{X},\mathbf{C}}}{1-R^2_{M \sim U|r,\mathbf{X},\mathbf{C}}} df}$. We can obtain an

unbiased estimate of $\alpha_r$ by fitting a mediator model shown in equation (5). Given the estimate, $\sqrt{Var(\hat{\alpha}_r)}$ can be consistently estimated by computing the sample variance of $\hat{\alpha}_r$. Likewise, we can obtain the estimate of $\beta_{res,m}$ by fitting an outcome model as shown in equation (13). Given the estimate, $\sqrt{Var(\hat{\beta}_{res,m})}$ can be consistently estimated by computing the sample variance of $\hat{\beta}_{res,m}$. A proof is shown in Appendix D.

Note that the standard error of disparity reduction shown in equation (15) is the same as the sample standard deviation of the disparity reduction estimate $(\hat{\delta}_{res}(r))$ if and only if the two sensitivity parameters are zero, meaning that there is no confounding in the mediator-outcome relationship given the intersectional group and measured confounders.

Calculating the standard error for disparity remaining is more complex. Therefore, we use $\tau(r, 0) = \delta(r) + \zeta(0)$ to approximately calculate the standard error for disparity remaining as

$$\begin{aligned} Var(\hat{\zeta}(0)) &\approx Var(\hat{\tau}(r,0)) + Var(\hat{\delta}(r)) - 2Cov(\hat{\tau}(r,0), \hat{\delta}_{res}(r)) + \\ &\quad 2kE[\sqrt{Var(\hat{\beta}_{res,m})}]Cov(\hat{\tau}(r,0), \hat{\alpha}_r) + 2kE[\hat{\alpha}_r]Cov(\hat{\tau}(r,0), \sqrt{Var(\hat{\beta}_{res,m})}), \end{aligned} \tag{16}$$

where $k = \sqrt{\frac{R^2_{Y \sim U|r,\mathbf{X},M,\mathbf{C}} \times R^2_{M \sim U|r,\mathbf{X},\mathbf{C}}}{1 - R^2_{M \sim U|r,\mathbf{X},\mathbf{C}}}} df$. Here, the estimates of $Var(\hat{\tau}(r,0))$ and

$Cov(\hat{\tau}(r,0), \hat{\delta}_{res}(r))$ can be obtained from sample variance and covariance matrix of the

initial disparity and disparity reduction estimates; $Cov(\hat{\tau}(r,0), \hat{\alpha}_r)$ can be obtained from sample covariance of the initial disparity and the regression coefficient $\hat{\alpha}_r$;

$Cov(\hat{\tau}(r,0), \sqrt{Var(\hat{\beta}_{res,m})})$ can be obtained from sample covariance of the initial disparity

and the standard error of $\hat{\beta}_{res,m}$.

Again, the standard error of disparity remaining shown in equation (16) is the same as the sample standard deviation of the remaining disparity estimate $(\hat{\zeta}_{res}(0))$ if and only if the two sensitivity parameters are zero, meaning that there is no omitted confounding existing in the mediator and outcome relationship.

This bias formulas in equation (14) and standard errors for disparity reduction and remaining in equations (15) and (16) can be used as sensitivity analysis that depends on two sensitivity parameters $R^2_{Y \sim U|r,\mathbf{X},M,\mathbf{C}}$ and $R^2_{M \sim U|r,\mathbf{X},\mathbf{C}}$. The two sensitivity parameters imply the degree to which an unobserved confounder is associated with the mediator and the outcome for the comparison group $R = r$ after conditioning on appropriate controls, expressed in $R^2$ values. Larger $R^2$ values of the sensitivity parameters indicate a larger bias for disparity reduction and remaining due to unobserved confounder $U$.

## 6.2. When the Linearity Assumption Is Violated

The expression of point estimates and standard errors becomes more cumbersome

when the interaction effect exists in the group-mediator relationships in the outcome

model. In such a case, the bias for disparity reduction and remaining estimates is the same

as equation (14), except that $\sqrt{Var(\hat{\beta}_{res,m})}$ should be replaced with

$\sqrt{\{Var(\hat{\beta}_{res,m}) + Var(\hat{\beta}_{res,rm}) + 2Cov(\hat{\beta}_{res,m}, \hat{\beta}_{res,rm})\}}$. For calculating standard errors for the disparity reduction and remaining, $\beta^2_{res,m}$ should be replaced with $(\beta_{res,m} + \beta_{res,rm})^2$.

Perhaps, a more straightforward way to address differential effects is to use an algorithm to change the reference group to $R = r$ when computing the effect of the mediator on the outcome ($\beta_{res,m}$) and its corresponding standard error for each comparison group ($\sqrt{Var}(\hat{\beta}_{res,m})$). For example, we fit the outcome model with the group-mediator interaction effect, setting the reference group to Black women. Then, the mediator effect on the outcome and its corresponding standard error for Black women can be obtained respectively as $\hat{\beta}_{res,m}$ and $\sqrt{Var(\hat{\beta}_{res,m})}$ in the outcome model even after adding the interaction effect. This algorithm provides a convenient and flexible way to address differential effects of the mediator or intermediate confounders by group status.

However, this algorithm does not address other nonlinear effects, such as the interaction effect in the mediator and intermediate confounder relationship. In this case, one should consider using the general bias formulas directly.

### 6.3. Robustness Values

Despite the development of numerous sensitivity analyses, not many social or medical scientists have used sensitivity analysis to assess the robustness of their findings. Recent literature on sensitivity analysis (Ding & VanderWeele, 2016; Cinelli & Hazlett, 2020) emphasized the use of standard way of reporting the robustness of research findings, which is expected to facilitate the discussions regarding how credible the estimated effect is to possible violations of no omitted confounding. For example, Ding and VanderWeele (2016) advanced the *E-value* that reports the robustness of research findings measured in the risk ratio; Cinelli and Hazlett (2020) advanced the robustness value (RV) that reports the robustness of research findings derived from linear regressions. In this section, we extend the RV computed for the treatment effect to disparity reduction and remaining.

We define the RV as the strength of association that will explain away the estimated disparity reduction or remaining, assuming an equal association to the mediator and the outcome, as $R^2_{Y \sim U|r,\mathbf{X},M,\mathbf{C}} = R^2_{M \sim U|r,\mathbf{X},\mathbf{C}} = \text{RV}$. Then, the RV for disparity reduction and remaining are respectively given by

$$
\begin{aligned}
\text{RV}_{\delta(r)} = &\frac{1}{2}\left(\sqrt{g_{\delta(r)}^4 + 4g_{\delta(r)}^2} - g_{\delta(r)}^2\right) \quad \text{and} \\
\text{RV}_{\zeta(0)} = &\frac{1}{2}\left(\sqrt{g_{\zeta(0)}^4 + 4g_{\zeta(0)}^2} - g_{\zeta(0)}^2\right)
\end{aligned}
\tag{17}
$$

where $g_{\delta(r)} = \frac{|\hat{\delta}_{res}(r)|}{|\alpha_r|\sqrt{Var(\hat{\beta}_{res,m})df}}$ and $g_{\zeta(0)} = \frac{|\hat{\zeta}_{res}(r)|}{|\alpha_r|\sqrt{Var(\hat{\beta}_{res,m})df}}$. A proof is given in Appendix E.

Next, we define $\text{RV}_{\alpha=0.05}$ as the strength of association that will change the significance of the estimated disparity reduction or remaining at the $\alpha = 0.05$ level, assuming an equal association to the mediator and the outcome. While the RV for disparity reduction and remaining can be computed easily from regression results, the $\text{RV}_\alpha$ cannot be computed easily. Therefore, we use a computational approach to obtain an approximate value of $\text{RV}_\alpha$. Specifically, we find combinations of two sensitivity parameters ($R^2_{Y \sim U|r,\mathbf{X},M,\mathbf{C}}$ and $R^2_{M \sim U|r,\mathbf{X},\mathbf{C}}$) that make the 95% confidence intervals (CI) of disparity reduction (or remaining) to cover approximately zero (i.e., $\delta(r) \pm t_{0.05,df}se(\delta(r)) < 0.001$). Once the combinations of two sensitivity parameters are identified that will make the CI approximately cover zero, we compute the average value of the two sensitivity parameters.

## 6.4. Illustration Using $R^2$-Based Sensitivity Analysis

Section 5.3 presents sensitivity analysis using regression coefficients. This section presents the same sensitivity analysis, but it is parameterized in $R^2$ values as defined in equation (14). Figure 3 presents the results for the sensitivity analysis of disparity reduction (A) and disparity remaining (B) based on two sensitivity parameters. The two sensitivity parameters are (1) the partial $R^2$ value of discrimination with CVH given the group status, mediator, and observed confounders, namely $R^2_{Y \sim U|r,\mathbf{X},M,\mathbf{C}}$ ($x$-axis), and (2) the partial $R^2$ value of discrimination with the education level given the group status and observed confounders, namely $R^2_{M \sim U|r,\mathbf{X},\mathbf{C}}$ ($y$-axis). We plot the points at which the estimated disparity reduction (remaining) becomes zero (bold line), and the 95% confidence intervals cover zero (dashed line) given the combination of two sensitivity parameters.

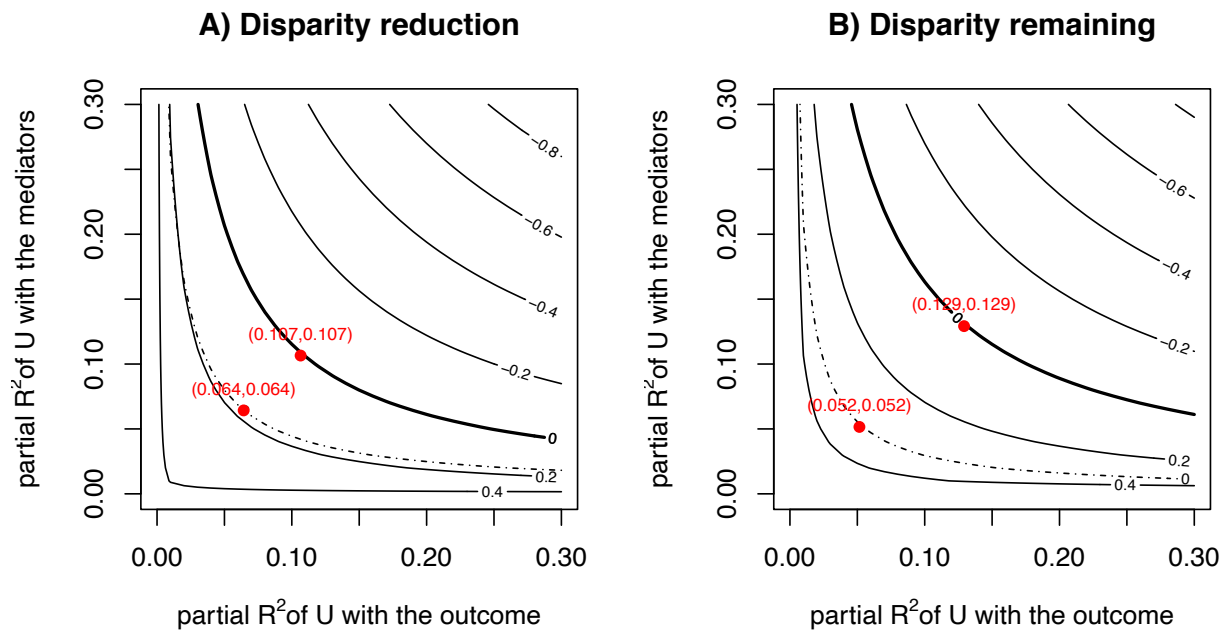

*Figure 3.* Sensitivity contour plots using $R^2$ values.

Note. 1) Bold lines represent the points at which the estimates become zero. 2) Standard lines represent the points at which the estimates become the respective value (e.g., -0.2, -0.1, 0.1, 0.2, etc.). 3) Dashed lines represent the points at which the upper and lower confidence intervals include zero. 4) Red points represents the robustness values (RVs), i.e., the partial $R^2$ values which make the estimates or upper/lower limits of confidence intervals zero, assuming equal $R^2$ values between the two sensitivity parameters.

The analysis indicates that the disparity reduction would still be negative (i.e., education significantly reduces the CVH gap between Black women and White men) if the unobserved confounder explains less than 10.7% of the variance of the mediator and the outcome after accounting for the existing confounders. The disparity reduction would still be significant at the 95% confidence level if the unobserved confounder explains less than 6.4% of the variance of the mediator and the outcome after accounting for the existing confounders.

Similarly, the disparity remaining would still be negative if the unobserved confounder explains less than 12.9% of the variance of the mediator and the outcome after accounting for the existing confounders, respectively. The disparity remaining would still be significant at the 95% confidence level if the unobserved confounder explains less than 5.2% of the variance of the mediator and the outcome after accounting for the existing confounders.

Although the qualitative classifications of effect size depend on the context of studies, we use Cohen (1988)'s guideline to judge how large the amount of confounding is. The amount of confounding required to change the disparity reduction (10.7%) and remaining (12.9%) estimates from negative to positive is considered medium. The amount of confounding required to change the significance of the disparity reduction (6.4%) and remaining (5.2%) estimates is considered small. These results indicate that the conclusion regarding disparity reduction and remaining due to the mediator (education) could be changed with an unobserved confounder that has a small effect on the mediator and the outcome after controlling for the existing confounders.

## 7. Discussion

In this study, we developed a set of sensitivity analyses that assesses the sensitivity of disparity reduction and remaining to possible violations of no-unobserved-mediator-outcome-confounding. Although we used the example of estimating disparity reduction and remaining after intervening on a mediator, the proposed sensitivity analyses can be used in any settings based on interventional indirect and direct effects.

Our study contributes to the fast-growing causal decomposition literature in several ways. First, we compared causal decomposition analysis based on interventional effects with causal mediation analysis based on natural effects. More importantly, we clarified that sensitivity analyses developed for natural indirect effects to possible pre-exposure confounding can only assess the sensitivity of disparity reduction when no intermediate confounding exists. We argued that the assumption (no intermediate confounding) is restrictive and may only be met when the mediator is measured shortly after the group status, for example, childhood poverty. Yet, such an assumption would be unrealistic in many studies that examine life-long health disparities between social groups.

Second, we derived general bias formulas for disparity reduction and remaining,

which serve as the basis of our proposed sensitivity analyses. The general bias formulas can be used beyond a particular statistical model and are applicable to any variable type of mediator, outcome, or intermediate confounder and do not require any assumptions. Moreover, the same bias formulas apply with pre-exposure and intermediate unobserved confounding. Within the causal mediation framework based on natural indirect effects, it has long been desired by many researchers to address both pre-exposure and intermediate unobserved confounding (Imai & Yamamoto, 2013; VanderWeele & Vansteelandt, 2014; Park & Esterling, 2020). Researchers can handle both types of confounding by utilizing the proposed sensitivity analyses, as the causal estimand is based on intervention direct and indirect effects. We acknowledge that not every situation is suitable for interventional direct and indirect effects. However, in some cases, such as the motivating example, defining interventional indirect effects makes more sense than natural indirect effects.

Third, we proposed a sensitivity analysis based on regression coefficients and $R^2$ values. The regression-based sensitivity analysis provides a straightforward way to assess the sensitivity of the effect estimates even without a specific software program. Compared to Park et al. (2020), there are two advantages: 1) the sensitivity analysis does not require a restrictive conditional independence assumption, and 2) sensitivity parameters are not conditional on the descendent of the exposure (group status). Although it hinges on the linearity assumption, we provided extensions that relax this assumption. We reparameterized regression-based sensitivity analysis to the scale-free $R^2$ values. The $R^2$-based sensitivity analysis is particularly useful for evaluating the sensitivity of the estimates' statistical inferences. In addition, robustness values provide a standard way to compare our findings' sensitivity to that of other studies.

We also acknowledge the limitations of the proposed sensitivity analyses. First, the $R^2$-based sensitivity analysis is only available for continuous outcomes. An extension to discrete outcomes is left for future research. Second, the proposed sensitivity analysis addresses unobserved confounding fixed in time. Addressing unobserved time-varying confounding would be an important generalization of this research. The time-varying confounding issue cannot be easily resolved within the causal mediation framework based on natural indirect effects, as VanderWeele et al. (2014) have pointed out. An alternative would be to use interventional indirect effects. Third, in addition to addressing unobserved confounders, it is crucial to address measurement errors in a mediator. A sensitivity analysis that simultaneously addresses unobserved confounders and measurement errors would be particularly beneficial to researchers investigating the role of psycho-social factors, which are susceptible to measurement errors, in reducing disparities.

## Appendix A: General Bias Formulas of $\delta(r)$ and $\zeta(0)$

The bias for disparity reduction for $r$ (bias($\delta(r)$)) is defined as the difference between the expected estimate and the true value. The bias equals

$$E[Y|r,\mathbf{c}] - \sum_{\mathbf{x},m} E[Y|r,\mathbf{x},m,\mathbf{c}]P(\mathbf{x}|r,\mathbf{c})P(m|R=0,\mathbf{c})-$$

$$E[Y|r,\mathbf{c}] + \sum_{\mathbf{x},m,u} E[Y|r,\mathbf{x},m,\mathbf{c},u]P(\mathbf{x},u|r,\mathbf{c})P(m|R=0,\mathbf{c})$$

$$= -\sum_{\mathbf{x},m} E[Y|r,\mathbf{x},m,\mathbf{c}]P(\mathbf{x}|r,\mathbf{c})P(m|R=0,\mathbf{c}) + \sum_{\mathbf{x},m,u} E[Y|r,\mathbf{x},m,\mathbf{c},u]P(\mathbf{x}|r,\mathbf{c},u)P(m|R=0,\mathbf{c})P(u|r,\mathbf{c})$$

$$= -\sum_{\mathbf{x},m,u} E[Y|r,\mathbf{x},m,\mathbf{c},u]P(u|r,\mathbf{x},m,\mathbf{c})P(\mathbf{x}|r,\mathbf{c})P(m|R=0,\mathbf{c})$$

$$+ \sum_{\mathbf{x},m,u} E[Y|r,\mathbf{x},m,\mathbf{c},u]P(\mathbf{x}|r,\mathbf{c},u)P(m|R=0,\mathbf{c})P(u|r,\mathbf{c})$$

$$= -\sum_{\mathbf{x},m,u} E[Y|r,\mathbf{x},m,\mathbf{c},u]P(u|r,\mathbf{x},m,\mathbf{c})P(\mathbf{x}|r,\mathbf{c})P(m|R=0,\mathbf{c})$$

$$+ \sum_{\mathbf{x},m,u} E[Y|r,\mathbf{x},m,\mathbf{c},u]\frac{P(u|r,\mathbf{x},c)}{P(u|r,c)}P(\mathbf{x}|r,\mathbf{c})P(m|R=0,\mathbf{c})P(u|r,\mathbf{c})$$

$$= \sum_{\mathbf{x},m,u} E[Y|r,\mathbf{x},m,\mathbf{c},u]\{P(u|r,\mathbf{x},c) - P(u|r,\mathbf{x},m,\mathbf{c})\}P(\mathbf{x}|r,\mathbf{c})P(m|R=0,\mathbf{c})$$

$$\tag{18}$$

The second equality holds because of the law of total probability. The third equality is due to Bayes' theorem.

The last expression of equation (18) is the general bias formula for $\delta(r)$. Since the bias for $\tau(r,0)$ due to omitted variable $U$ is zero, the bias for $\zeta(0) = -bias(\delta(r))$. This completes the proof.

## Appendix B: Sensitivity Analysis Using Regression Coefficients

Suppose that the following regression models for $Y$ and $U$ are correctly specified as equations (10). Then the bias for disparity reduction is given by

$$
= \sum_{\mathbf{x},m,u} E[Y|r,\mathbf{x},m,\mathbf{c},u]\{P(u|r,\mathbf{x},c)-P(u|r,\mathbf{x},m,\mathbf{c})\}P(\mathbf{x}|r,\mathbf{c})P(m|R=0,\mathbf{c})
$$

$$
= \sum_{\mathbf{x},m,u} E[Y|r,\mathbf{x},m,\mathbf{c},u]\{\sum_m P(u|r,\mathbf{x},m,c)P(m|r,\mathbf{x},c)\}P(\mathbf{x}|r,\mathbf{c})P(m|R=0,\mathbf{c})
$$

$$
- \sum_{\mathbf{x},m,u} E[Y|r,\mathbf{x},m,\mathbf{c},u]P(u|r,\mathbf{x},m,\mathbf{c})P(\mathbf{x}|r,\mathbf{c})P(m|R=0,\mathbf{c})
$$

$$
= \beta + \beta_r + \beta_x E[X|r,c] + \beta_m E[M|R=0,c] + \beta_c c + \beta_u\{\delta + \delta_r + \delta_x E[X|r,c] + \delta_m(\alpha + \alpha_r + \alpha_c c) + \delta_c c\}
$$

$$
- \Big[\beta + \beta_r + \beta_x E[X|r,c] + \beta_m E[M|R=0,c] + \beta_c c + \beta_u\{\delta + \delta_r + \delta_x E[X|r,c] + \delta_m(\alpha + \alpha_c c) + \delta_c c\}\Big]
$$

$$
= \beta_u \delta_m \alpha_r
$$

$$
(19)
$$

This completes the proof.

### Appendix C: Bias Formulas Based on the Coefficients of Determination

This proof is a straightforward extension of Cinelli and Hazlett (2020). Using equation (11), the bias for disparity reduction is given by

$$
\begin{aligned}
bias(\delta(r)) =& \alpha_r \beta_u \delta_m \\
=& \alpha_r \times \frac{Cov(Y,U|r,\mathbf{X},M,\mathbf{C})}{Var(U|r,\mathbf{X},D,M,\mathbf{C})} \times \frac{Cov(U,M|r,\mathbf{X},\mathbf{C})}{Var(M|r,\mathbf{X},\mathbf{C})} \\
=& \alpha_r \times \frac{Cor(Y,U|r,\mathbf{X},M,\mathbf{C})\sqrt{Var(Y|r,\mathbf{X},M,\mathbf{C})}}{\sqrt{Var(U|r,\mathbf{X},M,\mathbf{C})}} \\
& \times \frac{Cor(U,M|r,\mathbf{X},\mathbf{C})\sqrt{Var(U|r,\mathbf{X},\mathbf{C})}}{\sqrt{Var(M|r,\mathbf{X},\mathbf{C})}}.
\end{aligned}
\tag{20}
$$

where Cov=covariance and Cor=correlation. The third equality is because $Cov(A,B|C) = Cor(A,B|C)\sqrt{Var(A|C)}\sqrt{Var(B|C)}$.

Suppose that the partial $R^2$ value of unmeasured confounder $U$ with the outcome for $R = r$ given $\mathbf{X}$, $M$, and $\mathbf{C}$ be denoted as $R^2_{Y \sim U|r,\mathbf{X},M,\mathbf{C}}$; and suppose also that the partial $R^2$ value of unmeasured confounder $U$ with the mediators for $R = r$ given $\mathbf{X}$ and $\mathbf{C}$ be denoted as $R^2_{M \sim U|r,\mathbf{X},\mathbf{C}}$. Then, given equations (5) and (13), the absolute value of the bias can be expressed as

$$
\begin{aligned}
|bias(\delta(r))| =& |\alpha_r| \sqrt{\frac{R^2_{Y \sim U|r,\mathbf{X},M,\mathbf{C}} \times R^2_{M \sim U|r,\mathbf{X},\mathbf{C}}}{1 - R^2_{M \sim U|r,\mathbf{X},\mathbf{C}}}} \times \frac{\sqrt{Var(Y|r,\mathbf{X},M,\mathbf{C})}}{\sqrt{Var(M|r,\mathbf{X},\mathbf{C})}} \\
=& |\alpha_r| \sqrt{Var(\hat{\beta}_{res,m})} \sqrt{\frac{R^2_{Y \sim U|r,\mathbf{X},M,\mathbf{C}} \times R^2_{M \sim U|r,\mathbf{X},\mathbf{C}}}{1 - R^2_{M \sim U|r,\mathbf{X},\mathbf{C}}}} df
\end{aligned}
\tag{21}
$$

The first equality is derived from $\frac{Var(U|r,\mathbf{X},\mathbf{C})}{Var(U|r,\mathbf{X},M,\mathbf{C})} = \frac{1}{1 - R^2_{U \sim M|r,\mathbf{X},\mathbf{C}}} = \frac{1}{1 - R^2_{M \sim U|r,\mathbf{X},\mathbf{C}}}$ and $Cor^2(A,B|C) = R^2_{A \sim B|C}$. The second equality holds because $\sqrt{Var(\hat{\beta}_{res,m})} = \frac{\sqrt{Var(Y|r,\mathbf{X},M,\mathbf{C})}}{\sqrt{Var(M|r,\mathbf{X},\mathbf{C})}} \sqrt{\frac{1}{df}}$, where $\beta_m$ and $df$ are obtained from the outcome model shown in equation (13). This completes the proof.

**Appendix D: Standard Errors of Disparity Reduction and Remaining**

We first calculate the standard error for disparity reduction. The standard errors for $\hat{\beta}_{res,m}$ and $\hat{\beta}_m$ can be obtained respectively as

$$\sqrt{Var(\hat{\beta}_{res,m})} = \frac{\sqrt{Var(Y|r,\mathbf{X},M,C)}}{\sqrt{Var(M|r,\mathbf{X},C)}} \sqrt{\frac{1}{df}}$$

$$\sqrt{Var(\hat{\beta}_m)} = \frac{\sqrt{Var(Y|r,\mathbf{X},M,C,U)}}{\sqrt{Var(M|r,\mathbf{X},C,U)}} \sqrt{\frac{1}{df-1}} \tag{22}$$

The ratio of these standard errors is

$$\frac{\sqrt{Var(\hat{\beta}_m)}}{\sqrt{Var(\hat{\beta}_{res,m})}} = \frac{\sqrt{(1-R^2_{Y\sim U|r,\mathbf{X},M,C})}}{\sqrt{(1-R^2_{M\sim U||r,\mathbf{X},C})}} \sqrt{\frac{df}{df-1}}, \tag{23}$$

because $\sqrt{(1-R^2_{Y\sim U|r,\mathbf{X},M,C})} = \frac{\sqrt{Var(Y|r,\mathbf{X},M,C,U)}}{\sqrt{Var(Y|r,\mathbf{X},M,C)}}$ and $\sqrt{(1-R^2_{M\sim U||r,\mathbf{X},C})} = \frac{\sqrt{Var(M|r,\mathbf{X},C,U)}}{\sqrt{Var(M|r,\mathbf{X},C)}}$.
Therefore, we have $Var(\hat{\beta}_m) = Var(\hat{\beta}_{res,m})(\frac{1-R^2_{Y\sim U|r,\mathbf{X},M,\mathbf{C}}}{1-R^2_{M\sim U|r,\mathbf{X},\mathbf{C}}} \frac{df}{df-1})$.

Using the parametric identification result shown in Section 3.2, the standard error for disparity reduction for $R = r$ can be calculated approximately using the Delta method (Sobel, 1982) as

$$Var(\hat{\delta}(r)) \approx \alpha_r^2 Var(\hat{\beta}_m) + \beta_m^2 Var(\hat{\alpha}_r), \tag{24}$$

where $\beta_m = \beta_{res,m} - \sqrt{Var(\hat{\beta}_{res,m})} \sqrt{\frac{R^2_{Y\sim U|r,\mathbf{X},M,\mathbf{C}} \times R^2_{M\sim U|r,\mathbf{X},\mathbf{C}}}{1-R^2_{M\sim U|r,\mathbf{X},\mathbf{C}}}} df$ and
$Var(\hat{\beta}_m) = Var(\hat{\beta}_{res,m})(\frac{1-R^2_{Y\sim U|r,\mathbf{X},M,\mathbf{C}}}{1-R^2_{M\sim U|r,\mathbf{X},\mathbf{C}}} \frac{df}{df-1})$.

Next, we calculate the standard error for disparity remaining. Using

$\tau(r, 0) = \delta(r) + \zeta(0)$, we have

$$
\begin{aligned}
Var(\hat{\zeta}(0)) =& Var(\hat{\tau}(r, 0) - \hat{\delta}(r)) \\
=& Var(\hat{\tau}(r, 0)) + Var(\hat{\delta}(r)) - 2Cov(\hat{\tau}(r, 0), \hat{\delta}(r)) \\
=& Var(\hat{\tau}(r, 0)) + Var(\hat{\delta}(r)) - 2Cov(\hat{\tau}(r, 0), \hat{\delta}_{res}(r)) + 2Cov(\hat{\tau}(r, 0), bias(\delta(r))) \\
\approx& Var(\hat{\tau}(r, 0)) + Var(\hat{\delta}(r)) - 2Cov(\hat{\tau}(r, 0), \hat{\delta}_{res}(r)) \\
&+ 2kE[\sqrt{Var(\hat{\beta}_{res,m})}]Cov(\hat{\tau}(r, 0), \hat{\alpha}_r) + 2kE[\hat{\alpha}_r]Cov(\hat{\tau}(r, 0), \sqrt{Var(\hat{\beta}_{res,m})}),
\end{aligned}
$$
$$(25)$$

where $k = \sqrt{\frac{R^2_{Y \sim U|r, \mathbf{X}, M, \mathbf{C}} R^2_{M \sim U|r, \mathbf{X}, \mathbf{C}}}{1 - R^2_{M \sim U|r, \mathbf{X}, \mathbf{C}}} df}$. The third equality holds because

$\hat{\delta}(r) = \hat{\delta}_{res}(r) - bias(\delta(r))$. The fourth equality follows Goodman (1960) and uses the

conventional asymptotic approximation procedure (Stuart & Kendall, 1963). This completes the proof.

**Appendix E: Robustness Values for Disparity Reduction and Remaining**

We define the RV as the strength of association that will explain away the estimated disparity reduction (or remaining), assuming an equal association to the mediator and the outcome, as $R^2_{Y \sim U|r, \mathbf{X}, M, \mathbf{C}} = R^2_{M \sim U|r, \mathbf{X}, \mathbf{C}} = \text{RV}$. We find the RV such that $|bias(\delta(r))| = |\hat{\delta}(r)|$. Using equation 14, we have

$$|\alpha_r| \sqrt{Var(\hat{\beta}_{res,m})} \sqrt{\frac{\text{RV}^2_{\delta(r)}}{1 - \text{RV}_{\delta(r)}} df} = |\hat{\delta}(r)| \qquad (26)$$

Dividing both sides by $|\alpha_r| \sqrt{Var(\hat{\beta}_{res,m})} \sqrt{df}$, we have

$$\sqrt{\frac{\text{RV}^2_{\delta(r)}}{1 - \text{RV}_{\delta(r)}}} = \frac{|\hat{\delta}(r)|}{|\alpha_r| \sqrt{Var(\hat{\beta}_{res,m})} df} \qquad (27)$$

We define $g_{\delta(r)} \equiv \frac{|\hat{\delta}(r)|}{|\alpha_r| \sqrt{Var(\hat{\beta}_{res,m})} df}$ and solve for $\text{RV}_{\delta(r)}$. Then, $\text{RV}_{\delta(r)} = \frac{1}{2}(\sqrt{g^4_{\delta(r)} + 4g^2_{\delta(r)}} - g^2_{\delta(r)})$. The RV for $\zeta(0)$ can be obtained the same way and thus is omitted. This completes the proof.